\documentstyle[12pt,epsf]{article}
\begin{document}

\newcommand{\disp}{\displaystyle}

\newcommand{\be}{\begin{equation}}
\newcommand{\ee}{\end{equation}}
\newcommand{\ba}{\begin{array}}
\newcommand{\ea}{\end{array}}
\newcommand{\ben}{\begin{eqnarray}}
\newcommand{\een}{\end{eqnarray}}
\newcommand{\bc}{\begin{center}}
\newcommand{\ec}{\end{center}}
\newcommand{\eq}[1]{(\ref{#1})}
\newcommand{\tw}{t_{\rm w}}
\newcommand{\eql}[2]{\begin{equation}\label{#1} #2 \end{equation} }

\newcommand{\avg}[1]{\left\langle #1 \right\rangle }
\newcommand{\peq}{p^{\rm eq}}
\newcommand{\non}{\nonumber}
\newcommand{\invN}{\frac{1}{N}}

\title{Hierarchical Diffusion, Aging and Multifractality}

\author{Hajime Yoshino\thanks{Institute for Solid State Physics,
Univ. of Tokyo, 7-22-1 Roppongi, Misato-ku, 
Tokyo, 106 Japan. 
(E-mail: yhajime@ginnan.issp.u-tokyo.ac.jp)}\\
Institute of Physics, University of Tsukuba, Tsukuba, Japan}

\maketitle

\begin{abstract}
We study toy aging processes in hierarchically
decomposed phase spaces
where the equilibrium probability distributions are multifractal.
We found that the an auto-correlation function, survival-return
probability, 
shows crossover behavior from a power law $t^{-x}$ in the quasi-equilibrium 
regime ($t\ll\tw$)
to another power law $t^{-\lambda}$ ($\lambda \geq x$) 
in the off-equilibrium regime ($t\gg\tw$)
obeying a simple $t/\tw$ scaling law.
The exponents $x$ and $\lambda$ are related with the
so called {\it mass exponents}
which characterize the multifractality.
\end{abstract}

\newpage

\section{Introduction}

The {\it aging} processes, i.~e. relaxational processes to approach 
the thermal equilibrium,  are extremely slow 
in glassy systems like spin-glasses and one can observe 
remarkable {\it aging effects} in experiments\cite{RPRM,STRUIK}. 
Among the various phenomenological descriptions of the aging effects
are due to Sibani and Hoffman \cite{SH} who proposed a scenario
based on the concept, {\it hierarchical diffusion} \cite{HS88}.
The latter concept has been implemented in many toy models 
\cite{HK85}-\cite{BH}, which we hereafter refer to as {\it hierarchical
diffusion models}. The concept \cite{Pal} is roughly as the following.
Firstly, one considers that the free-energy landscape 
consists of hierarchically nested valleys, which are usually 
described in terms of a certain {\it tree} structure. 
Then one introduces a relaxational dynamics in terms of a certain master 
equation which describes diffusion processes between different valleys driven 
by thermal hoppings over the barriers. Solving the master equation, one 
obtains the time evolution of the distribution of probabilities to find the 
system at different bottoms of valleys (or {\it leaves} of the trees). 

In the present paper, we consider hierarchical diffusion in a class of trees 
which have the following two characteristics. Firstly, we consider that the
backbone structures of the trees
have self-similarity as in many of the previously studied models. Secondly, 
we consider that the equilibrium probability distributions on the leaves
are multifractal. The latter point is different from the previously 
studied models, whose equilibrium probability distributions are restricted 
to be uniform by their designs ( see however \cite{Nemo88}).
We introduce the relaxational dynamics in terms of an exactly solvable 
master equation.

We study aging processes after rapid temperature quenches in our model.
An aging process appears as the growth of 
sub-trees in which the probability distributions are quasi-equilibrium 
(multifractal) whereas on larger scales than such sub-trees 
the probability distributions are still non-equilibrium (non-multifractal).
As the result, a simple auto-correlation function,
the survival-return probability,  shows a characteristic
crossover behavior: it decays by a power law $t^{-x}$ 
in the quasi-equilibrium regime ($t\ll\tw$) but by
another power law $t^{-\lambda}$ in the off-equilibrium regime ($t\gg\tw$)
and obeys a simple $t/\tw$ scaling.
The exponents $x$ and $\lambda$ turns out to be related with the so called
{\it mass} exponents which characterize the multifractal properties of 
the probability distributions. 

The organization of this paper is the following. In section 2, we introduce
our hierarchical diffusion model. In section 3, we analyze the aging effect 
in our model focusing on the scaling properties of an autocorrelation 
function. In section 4, we summarize this paper with some discussions.

\section{The Model}

\subsection{Construction of Tree}

Let us consider a system of many {\it states} 
which have the following clustering property.
Suppose that the system can be coarse-grained
so that different {\it states} merge into fewer number of  {\it states}.
In Fig.~1, we show such a system represented as a tree.
The magnitude of the resolution power increases downward along the $h$ 
axis: the {\it states} (branches) are differentiated
into more states (branches) as we raise the resolution power.
We label the {\it states} differentiated with the maximum resolution power
( the {\it leaves} on the baselines of the trees) 
as $L_{i}$ ($i=1\ldots N$) where $N$ is the number of such states. 

For simplicity, we will only consider 
bifurcating trees: at every branch point we always have two 
branches stemming down.
We will refer to the set of branches and branch points under 
a branch point, say $A$, as sub-tree $A$. 
We will refer one and the other side under a branch point $A$ 
as side-{\it I} and {\it II} of $A$ and present them on the left
and right side respectively in the figures below.

We construct the partition function of the system at the equilibrium as the 
following. The coarse-grained states (branches) 
at the same resolution level are considered to be all
energetically degenerate with each other. However, the
number of {\it microscopic} states they contain can vary, which means that
the equilibrium local entropies and so the equilibrium
statistical weights can vary as well.
We consider that as a bifurcation takes place downward along the $h$ axis
(the direction to increase the resolution power),
the equilibrium local entropy of a coarse-grained state is partitioned into 
those of two sub-states underneath. In order words,
the partition function of sub-tree B is partitioned into those of
side-$I$ and side-$II$ sub-trees 
under $B$ with a certain partition ratio, say $1-\psi(B)$ and $\psi(B)$
($0 \leq \psi \leq 1$), respectively.

Let us denote the child of $B$ (a branch point just below $B$)
on side-$I$ as $C_{I}(B)$ and side-$II$ as $C_{II}(B)$.
Then the ratio of the partition function of sub-tree 
$C_{I}(B)$ and $C_{II}(B)$ to that of $B$, 
which we denote as $\pi(\acute{B},B)$,
takes the following values,
\eql{rw}{\pi(\acute{B},B)=\left\{\begin{array}{cl}
1-\psi_{B} & \mbox{$\acute{B}=C_{I}(B)$}\\
\psi_{B}   & \mbox{$\acute{B}=C_{II}(B)$}
\end{array} \right..}
It is useful to generalize the above argument as the following.
Suppose that a sub-tree $D$ is enclosed by a larger sub-tree $B$.
Let us denote the {\it parent} of $D$ 
( the branch point just above $D$) 
as $D_{1}$ and the {\it grand-parent} of $D$ ( the {\it parent} of $D_{1}$) 
as $D_{2}$ and so on. 
Suppose that  $B$ is the $K$'th {\it ancestor} of $D$, i.~e.~$D_{K}=B$.
Then the ratio of the partition function 
of sub-tree $D$ to that of sub-tree $B$,
which we denote as $\pi(D,B)$,
can be written as the product of $\pi$'s 
along the (unique) vertical {\it path} which connects $D$ and $B$,
\eql{prod_pi}{\pi(D,B) \equiv
\pi(D,D_{1})\pi(D_{1},D_{2})\ldots\pi(D_{K-1},B).}
For example, consider the set of leaves 
in a sub-tree $B$.
Then $\pi(L_{i},B)$ associated with such a leaf $L_{i}$ 
can be interpreted as the {\it relative} statistical weight of 
the leaf among the set of leaves in sub-tree $B$.
We denote the highest branch point as $B_{\rm top}$ and 
choose the partition function of the whole tree to be $1$.
Then the equilibrium statistical weight of a leaf $L_{i}$ can be
written as $\pi(L_{i},B_{\rm top})$.

Here we define some other terminologies for later uses.
We denote the {\it ancestors} and {\it descendants}, which
are the set of branch points above and below the branch point $B$, 
as ${\cal A}(B)$ and ${\cal D}(B)$ respectively. 
We denote the set of all branch points under
side-$I$  and $II$ of $B$ as ${\cal D}_{I}(B)$ 
and ${\cal D}_{II}(B)$ respectively.
(Note that  ${\cal D}_{I}(B) \cup {\cal D}_{II}(B)={\cal D}(B)$.)
We denote the {\it parent}, {\it grand-parent}
and  the $K$'th ancestor of $B$ as $B_{1}$, 
$B_{2}$ and $B_{K}$ respectively as we already did above.
We denote the {\it lowest common ancestor} of $B$ and $\acute{B}$,
the branch point at the top of the smallest sub-tree which
enclose both $B$ and $\acute{B}$, as $A(B,\acute{B})$.
For the convenience, we also introduce a hypothetical branch point 
$B_{\rm ceiling}$ whose height is $h_{B_{\rm ceiling}}=\infty$
and set $\pi(B_{\rm top},B_{\rm ceiling})=1$.

\subsection{The Master Equation}

We now introduce a stochastic dynamics of the temporal state 
in the hierarchically decomposed phase space. 
We consider that there is an thermally activated {\it excitation}
associated with a branch point, say $B$, in the tree 
with which the temporal state can go from one to the other leaves in
sub-tree $B$.
The excitations associated with the branch points at higher $h$ 
are considered to have higher activation energies to be excited. 
Thus we now redefine the vertical axis $h$ as the scale of 
the activation energies of such excitations associated with
the branch points.

We now introduce a simple exactly solvable dynamics which describes
the stochastic jumps between 
the leaves, i.~e. the states which are differentiable
with the maximum resolution power.
Here after we call the latter simply as {\it states}.
Let us denote the probability to find the temporal state of the
system at a state $L_{i}$
at time $t$ as $p_{i}(t)$. The time dependent distribution
of the probability can be expressed in terms of  a vector,
\eql{pt}{{\bf p}(t)=(p_{1}(t),p_{2}(t),\ldots,p_{N}(t)),} 
which should become equal to the equilibrium probability distribution
in the limit $t \rightarrow \infty$. We denote the latter as 
 as ${\bf p}^{\rm eq}=
(\peq_{1},\peq_{2},\ldots,\peq_{N})$  where 
\eql{peq}{\peq_{i}\equiv\pi(L_{i},B_{\rm top}).}

Let us denote  the transition probability 
to go from  state $L_{j}$ to $L_{i}$ 
in a unit time as ${\bf W}_{ij}$. 
Then the master equation for the evolution of the probabilities 
can be written as
\eql{Meq}{\disp\frac{d}{dt}{\bf p}(t) = -{\bf \Gamma} {\bf p}(t),}
with
\eql{Gamma}{-{\bf \Gamma}_{ij}={\bf W}_{ij}-\delta_{ij}
\sum_{k}{\bf W}_{kj}.}
Note that the sum of the probability $\sum_{i=1}^{N} p_{i}(t)$
is always conserved.

We consider that the thermal jump process of the temporal state from a 
state (leaf) $L$ to another state consists of two stages. 
In the first stage, the excitations associated with the branch points 
in ${\cal A}(L)$ (ancestors of $L$) are activated in a successive manner
as the following.
Suppose that the excitation associated with such a branch point $B$
is activated. Then its parent $B_{1}$ get a chance to become active 
or not with the probability $\exp(-(h_{B_{1}}-h_{B}))$ and
$1-\exp(-(h_{B_{1}}-h_{B}))$ respectively.
If $B_{1}$ becomes active, we repeat the same trial for $B_{2}$.
Otherwise, the successive ignition of the excitations 
stop there at the level $h_{B}$. Thus the probability
that the 1st stage ends at $B$ is 
\ben
w(B\leftarrow L)&=&\prod_{n=1}^{K} \exp(-(h_{A_{n}}-h_{A_{n-1}}))
\left(1-\exp(-(h_{B_{1}}-h_{B}))\right)\non\\
&=&\exp(-h_{B})-\exp(-h_{B_{1}})
\een
where $A_{0}=L$ and $A_{K}=B$.

The 2nd stage is the 
{\it falling down} process from the height $h_{B}$ to a leaf of sub-tree $B$.
Reminding that the leaves $\acute{L} \in {\cal D}(B)$  have different relative 
statistical weights $\pi(\acute{L},B)$, 
we expect that the probabilities to fall into the leaves
depend on their amount of local entropies 
in such a way that those with larger amount of local entropies 
have more chances to {\it receive} the temporal state. 
So we simply choose the probability to fall down to
a state (leaf) $\acute{L}$ as 
\be
w(\acute{L} \leftarrow B)=\pi(\acute{L},B).
\ee
Combining the above two factors, 
we obtain the transition 
probability to go from $L$ to $\acute{L}$ via $B$ as
\ben
w(\acute{L}|B|L)&=&w(\acute{L} \leftarrow B)w(B\leftarrow L)\non\\
&=&\pi(\acute{L},B)\left[\exp(-h_{B})-\exp(-h_{B_{1}}) \right] \label{via}.
\een
Note however that such a process takes place only if both $L$ and $\acute{L}$ 
belong to the sub-tree $B$.

The transition probability $W_{ij}$ from $L_{j}$ to $L_{i}$ is the sum of
the transition probabilities over $A_{n}$ of 
$n=0,1,\ldots,M-1$ where $A_{0}=A(L_{i},L_{j})$ (the lowest common ancestor
of $L_{i}$ and $L_{j}$)
and $A_{M}=B_{\rm ceiling}$. 
Thus the off-diagonal-elements of the matrix ${\bf W}$ becomes
\eql{Wij}{
{\bf W}_{i\neq j}= \sum_{n=0}^{M-1} w(L_{j}|A_{n}|L_{i})
=\disp\sum_{n=0}^{M-1}
\left[ \exp(-h_{A_{n}})-\exp(-h_{A_{n+1}}) \right]
\pi(L_{i},A_{n}), }

Using \eq{peq} and \eq{Wij}, it can be checked that 
the detailed balance condition
\eql{d_balance}{{\bf W}_{ij}\peq_{j}={\bf W}_{ji}\peq_{i},}
is satisfied with this choice. 

\subsection{Solution of the Master Equation}

We now solve the master equation \eq{Meq}.
The formal solution  can be written as
\eql{formal}{{\bf p}(t)=\exp(-{\bf \Gamma} t){\bf p}(0)}
where ${\bf p}(0)$ is the initial distribution at $t=0$.
The probability that the temporal state which initially stay
at $L_{j}$ reach $L_{i}$ at time $t$ is
\eql{G}{{\bf G}_{ij}(t)=\left[\exp(-{\bf \Gamma}t)\right]_{ij}.}
In order to calculate the propagator ${\bf G}$,
it is convenient to introduce a new matrix ${\bf \tilde{\Gamma}}$
with which we can rewrite ${\bf \Gamma}$ as 
\eql{tildeG}{{\bf \Gamma}\equiv({\bf p}^{\rm eq})^{1/2}\tilde{\bf \Gamma}
({\bf p}^{\rm eq})^{-1/2},}
where  ${\bf p}^{\rm eq}$ is the vector of equilibrium statistical
weights $\peq_{i}$ defined in \eq{peq}.
Note that $\tilde{\bf \Gamma}$ is a 
real symmetric matrix so that it has real eigen values.

We now look for the $N$ eigen states of $\tilde{\Gamma}$ in a heuristic way.
First of all, the {\it static} mode can be found easily as the following.
The matrix $\Gamma$ satisfies 
${\bf\Gamma}{\bf p}^{\rm eq}=0$ due to
\eq{Gamma} and  \eq{d_balance}, from which we obtain
$\tilde{\bf \Gamma}({\bf p}^{\rm eq})^{1/2}=0$. 
The last equation means that the vector $({\bf p}^{\rm eq})^{1/2}$
is an eigenvector whose eigenvalue is $0$ i.~e. {\it static} mode.
There are $N-1$ other eigen states ({\it dynamic modes}) left to be 
found.

We construct here a set of vectors which consists of 
vectors localized under the branch points.
We may call this set of vectors
as {\it umbrella set} because of the localized shape of the amplitudes.
On a branch point $B$ we define a vector
\eql{umb}{\hat{S}_{i}(B)=\pi^{1/2}(L_{i},B)u(L_{i},B)}
where
\eql{u_d}{ u(\acute{B},B)=\left \{\begin{array}{cll}
                  1          & B=B_{\rm ceiling}  & \\ 
\sqrt{\disp\frac{\psi_{B}}{1-\psi_{B}}} & B \neq B_{\rm ceiling}
                            & \acute{B} \in {\cal D}_{\rm I}(B)\\
-\sqrt{\disp\frac{1-\psi_{B}}{\psi_{B}}} & B \neq B_{\rm ceiling}
                             & \acute{B} \in {\cal D}_{\rm II}(B)\\
            0                 &   & \mbox{otherwise.}
\end{array} \right. }
Note that the vector $\hat{S}_{i}(B_{\rm ceiling})$ is
identical to the eigen vector of the static mode we obtained above.
It can be easily checked that the vectors are normalized and 
orthogonal,
\eql{orthogonal}{
\sum_{i}\hat{S}_{i}(B)\hat{S}_{i}(\acute{B})=\delta_{B,\acute{B}}.}
Since there are $N-1$ eigen vectors on $N-1$ branch points
and one static mode, which are all linearly independent with each other,
they together constitute a complete set of dimension $N$.
As shown in Appendix A, the vector $\hat{S}_{i}(B)$
actually turn out to be the correct 
eigen vectors of $\tilde{\bf \Gamma}$  
which have the associated eigen values
\eql{z_umb}{\hat{z}(B)=\exp(-h_{B}).} 
Note that $\hat{z}(B_{\rm ceiling})=0$ (static mode) is endured since 
we have set $h_{B_{\rm ceiling}}=\infty$. 

Now  we rewrite some previously defined  matrices in terms of the 
umbrella set.
At first, $\tilde{\bf \Gamma}$ becomes
\eql{tildeG_re}{\tilde{\bf \Gamma}_{ij}
=\sum_{B}\hat{S}_{i}(B)\hat{z}(B)\hat{S}_{j}(B).}
Then using the last equation and \eq{umb} in \eq{tildeG}, 
the matrix ${\bf \Gamma}_{ij}$  becomes
\ben
{\bf \Gamma}_{ij} &=&
(\peq_{i})^{1/2}\tilde{\bf \Gamma}_{ij}(\peq_{j})^{-1/2}\non\\
&=&\sum_{B}\pi(L_{i},B)u(L_{i},B)u(L_{j},B)\hat{z}(B)\label{Gamma_re}.
\een
Finally, we also rewrite 
the propagator ${\bf G}$ in terms of the 
umbrella set. Using \eq{G}, \eq{tildeG} and \eq{tildeG_re} we obtain,
\ben
{\bf G}_{ij}(t) 
&=& (\peq_{i})^{1/2}
\sum_{B} \left \{ \hat{S}_{i}(B)\exp(-\hat{z}(B)t)\hat{S}_{j}(B) \right \}
(\peq_{j})^{-1/2} \non\\
&=& \sum_{B}\pi(L_{i},B)u(L_{i},B)u(L_{j},B)\exp(-\hat{z}(B)t). \label{re_G}
\een
Using \eq{umb} and \eq{u_d} 
and performing similar calculus shown in the appendix, we obtain 
the propagator in a more explicit form,
\ben
{\bf G}_{ij}(t) 
&=&  \disp\sum_{n=0}^{M-1}
\left[ \exp(-\hat{z}(A_{n+1})t)
      -\exp(-\hat{z}(A_{n})t) \right]\pi(L_{i},A_{n})\non\\
&+& \delta_{ij}\exp(-\hat{z}(A_{0})t)
\een
where we defined $A_{0}=A(L_{i},L_{j})$ and $A_{1},A_{2},\ldots,A_{M-1}$,
$A_{M}=B_{\rm ceiling}$.


\subsection{Multifractality on Self-similar Trees}\label{mf}

\subsubsection{Mass Exponents}

We consider in this paper self-similar trees on which the distributions of 
the equilibrium statistical weights  have the following 
multifractal characteristics. 
Let us define the {\it $q$-th  moments} of the statistical weights 
\cite{multifractal} as,
\ben
M_{q}(h) dh & \equiv &
\disp\overline{\sum_{\acute{B} \in {\cal D}(B)}\delta(h-(h_{B}-h_{\acute{B}}))
\pi^{q}(\acute{B},B)} dh \label{Mq}
\een
where the over-line means 
the average over statistically independent sub-trees $B$. 
Suppose that a $q$-th moment have the following scaling behavior,
\eql{Mq_tauq}{M_{q}(h) dh \sim \exp(\tau(q)h) dh,}
where the exponent $\tau(q)$ is called as a {\it mass exponent}
\cite{multifractal}. If  $\tau(q)$ depends non-linearly 
on $q$, the distribution is regarded as
{\it multifractal}. 

The geometrical self-similarity appears in the 0th moment 
$M_{0}(h)$, which is just the number of leaves of a tree of
height $h$. The mass exponent $\tau(0)$ is
the fractal dimension of the tree and sometimes called
as {\it silhouette} of the tree\cite{BH},
which measures whether a given tree is {\it slender} or {\it fat}.
A trivial exponent is $\tau(1)$ which is always $0$ because of the
normalization condition of the statistical weights. 
Another exponent which turns out to be
quite important is the mass exponent of 
the 2nd moment  $\tau(2)$.
As we see later, the two exponents $\tau(0)$  and $\tau(2)$
are related with the dynamic exponents of the
survival-return probability in the present model.

\subsubsection{Randomly Branching Trees}\label{RBP}

\newcommand{\pbra}{p_{\rm branch}}

As an example of the trees which have the multifractal 
properties mentioned above,
we construct here a specific class of random trees
generated by the following randomly branching process (RBP). 
We use this specific example later when some demonstrations are necessary.

An RBP
starts with a single {\it leaf}, which is regarded as 
the highest branch point $B_{\rm top}$ and two branches stemming down from it.
In a single step, the length of
the branches under the lowest (new) branch points 
get longer by one unit, say $db$ so that the height (from the 
baseline of the tree) of 
all branch points and so that the height of the tree get
larger by $db$. 
The branching at a leaf
occurs with probability $\pbra$ in a single step. 
When it takes place, the leaf becomes 
a new branch point and  two new branches starts  from it.
Each of such events occurs independently from each other.
Repeating this procedure, we obtain a backbone structure of a tree.

Next we assign the {\it weights} on the tree.
Consider a branch point $B$ and its child (the branch points just below
B) $C_{I}$ on side-$I$ and $C_{II}$ on side-$II$.
We determine the variable $\psi_{B}$  and assign
$\pi(C_{I},B)=1-\psi_{B}$ and $\pi(C_{II},B)=\psi_{B}$ to the branches 
on the side-$I$ and $II$ of $B$. 
The variable $\psi_{B}$ on each branch point is chosen randomly 
from the distribution \eql{dist_prob}{F(\psi) d\psi
 \equiv \mbox{probability that $\psi_{B}$ lies between
$\psi$ and $\psi+d\psi$}.}
We perform this procedure for the whole branch points.

Consider a tree which grows larger by the RBP.
A natural consequence of the RBP is that the backbone structure
possesses statistical self-similarities.
On the other hand, the statistical weights on the leaves are 
successively partitioned further into more and more fine pieces by the RBP.
It is well known that if such a process is repeated, one oftenly finds very 
peculiar distribution of the weights : some set of pieces which have rather 
larger weights but negligibly smaller population compared with 
the {\it typical} ones 
come to rapidly dominate the total sum of the statistical weights 
(the partition function of the whole tree) as the branching proceed further.
This phenomena is called as {\it curdling}
\cite{Mandel} of  multifractal objects.

As we show in Appendix B, we actually obtain the scaling
property of the form \eq{Mq_tauq} in the case of the random trees
generated by the RBP. The mass exponent is,
\eql{tauq}{\disp\tau(q)=db^{-1}
\log \left [\ \pbra \disp\int_{0}^{1} d\psi F(\psi) 
\left\{\psi^{q}+(1-\psi)^{q}\right\}
+(1-\pbra) \right ].}
It can be seen that it is generally non-linear with $q$. 
In Fig.~2 we show an example of $\tau(q)$ 
on random trees generated by a RBP.
A special case when $\tau(q)$ becomes linear with $q$ is 
when the following two conditions hold: 
$p_{\rm branch}=1$ (deterministic branching) and 
$F(\psi)=\delta(\psi-1/2)$ (always symmetric partition).


The real samples of such random trees
can be generated numerically by 
by the following  Monte Carlo Method.
In one Monte Carlo step (MCS), the height of all branch 
points are raised by $db$. Simultaneously a pseudo random number is
generated for every leaf and if it is smaller than $\pbra$, 
a bifurcation takes place:
the leaf becomes a new branch point
and two new branches start from it.
Fig.~1 is actually an example of such a random tree obtained by 
simulating the
RBP of $\pbra=0.10$, $db=1.0$ and 
$F(\psi)=\delta(\psi-0.2)$ for $20$ MCS.

\section{Aging Effect}

\subsection{Growth of Quasi-Equilibrium Domain}

We now consider an aging process  after rapid temperature quench
from high temperature.
For this purpose, we 
choose the initial configuration as,
\eql{initial}{{\bf p}(0)=\frac{1}{N}.}

After waiting for $\tw$ (waiting time),
the probability distribution evolves up to,
\eql{ptw}{{\bf p}(\tw) = {\bf G}(\tw){\bf p}(0),}
which eventually become ${\bf p}^{\rm eq}$ as $\tw \rightarrow \infty$.
Note that the initial non-equilibrium distribution is not multifractal 
because it is uniform, 
while the final fully-equilibrated distribution is multifractal.
Hence the aging process in the present context can be understood
as the process to approach a multifractal distribution 
from a non-multifractal distribution.

In order to see how the system {\it age}, it is 
convenient to define
\eql{r_tw}{r_{i}(\tw) \equiv p_{i}(\tw)/\peq_{i}
= \invN\sum_{B}\exp(-\hat{z}(B)\tw)\pi^{-1}(B,B_{\rm top})
\tilde{u}(L_{i},B),}
where we used \eq{re_G} in \eq{ptw} and defined
\eql{tilde_u}{\tilde{u}(L_{i},B)=u(L_{i},B)
\sum_{j\in {\cal D}(B)}u(L_{j},B).}
In Fig.~3 we show an example of $r_{i}(\tw)$ calculated 
using the exact solution of the master equation solved 
on a real sample of random tree shown in Fig.~1.
One can see that, as $\tw$ increases, $r_{i}(\tw)$
of different states come to join with each other successively and constitute 
groups, among each of which the values of $r_{i}(\tw)$ are common.
Note that as far as the transitions within such groups are concerned,
the detailed balance \eq{d_balance} is fulfilled. 
Thus we may call such a group of states as 
a {\it quasi-equilibrium domain}. 

In order to understand the growth mechanism of the 
{\it quasi-equilibrium domain} in a more formal way,
let us introduce a characteristic height $h^{\rm eff}(\tw)$ 
which grows logarithmically with $\tw$,
\eql{htw}{h^{\rm eff}(\tw)\equiv \log(\tw).}
Due to the factor 
$\exp(-\hat{z}(B)\tw)\simeq\exp(-\exp(h^{\rm eff}(\tw)-h_{B}))$ , 
the contributions from the branch points lower than $h^{\rm eff}(\tw)$
in the r.h.s. of the equation \eq{r_tw} are negligibly small, 
compared with those from the branch points higher than 
$h^{\rm eff}(\tw)$. So it can be roughly approximated as
\eql{approx_r_tw}{
r_{i}(\tw)\simeq \sum_{h(B)\gg h^{\rm eff}(\tw)}
\exp(-\hat{z}(B)\tw)\pi^{-1}(B,B_{\rm top})
\tilde{u}(L_{i},B).}
Consider a pair of states $L_{i}$ and $L_{j}$
whose lowest common ancestor is $A(L_{i},L_{j})$.
Suppose that after certain waiting time $\tw^{*}$, the characteristic height 
$h^{\rm eff}(\tw^{*})$ becomes larger than $h_{A(L_{i},L_{j})}$.
Then from the definitions \eq{tilde_u} and \eq{u_d}, all the terms
that survive in the sum of \eq{approx_r_tw}
become the same for $L_{i}$ and $L_{j}$.
Consequently $r_{i}(\tw)= r_{j}(\tw)$ holds 
forever for $\tw \gg \tw^{*}$.

To summarize, the aging process of the present hierarchical model is 
understood as the growth of {\it aged} sub-trees or quasi-equilibrium
domains. Here we mean by  a quasi-equilibrium
domain a sub-tree under a branch point
lower than the characteristic height $h^{\rm eff}(\tw)$, which grows
logarithmically with $\tw$.
The probability distributions inside an aged sub-tree
is almost the same as that of fully-equilibrated 
one except for a common multiplicative factor.
This is one of the most important consequence
of {\it hierarchical diffusion}
and actually found to be the case in the relaxational 
dynamics of microscopic spin-glass models 
\cite{SS94}.

The probability distribution inside an aged sub-tree 
is multifractal  and possess the same scaling behaviors as
that of the fully-equilibrated one while 
on higher scale $h \gg h^{\rm eff}(\tw)$, the system is still 
highly non-equilibrium or {\it young} in the sense that the probability 
distribution is not multifractal.

\subsection{Survival-Return Probability}\label{srp}

We now introduce one of the simplest auto-correlation
function as a prove, that is the survival-return probability,
\eql{udiff_q}{q(t+\tw,\tw) \equiv 
\sum_{i}{\bf p}_{i}(\tw){\bf G}(t)_{ii}.}
It measures the probability that the system returns to the leaf
where it stayed at time $\tw$ after additional traveling of $t$.

Using \eq{re_G}, the autocorrelation function can be rewritten 
in terms of the umbrella set as
\ben
q(\tw,t+\tw)  & = & 
\sum_{i}{\bf p}_{i}(\tw){\bf G}(t)_{ii} \non\\
&=& \sum_{B}  \exp(-\hat{z}_{B}t)
\sum_{i} p_{i}(\tw)\pi(L_{i},B)u^{2}(L_{i},B)\non\\
& = & \int dz \rho_{\tw}(z)  \exp(-zt) \label{q_rho} ,
\een
where we defined a kernel $\rho_{\tw}(z)$ as
\be
 \rho_{\tw}(z) \equiv 
\sum_{B}  \delta(z-\hat{z}_{B})
\left \{ \sum_{i} p_{i}(\tw)\pi(L_{i},B)u^{2}(L_{i},B)
\right \}. \label{rho_tw}
\ee
In the latter subsections, we study the scaling behaviors of the
survival-return probability in random trees generated by RBP
focusing on the role played by the waiting time $\tw$.

\subsection{Two Extreme Cases}

\subsubsection{Zero Waiting Time}\label{zero}

We consider at first a special case of zero-waiting time case 
$\tw=0$, which means
that the system is in an extremely non-equilibrium condition
at $t=0$.
Since we have set the initial condition as  \eq{initial}, we obtain
\ben
\rho_{0}(z) & = & \invN\sum_{B}\delta(z-\hat{z}(B))\non\\
&=& \invN\sum_{B}\delta\left(h-h_{B})\right) dh, \label{rho_0}
\een
where we used \eq{z_umb} and defined a variable $h\equiv -\log(z)$.

Using \eq{Mq} and \eq{Mq_tauq}, we obtain
\ben
\invN\sum_{B}\delta(h-h_{B}) dh 
&=& \invN\sum_{B\in {\cal D}(B_{\rm top})}
\delta \left ((h_{B_{\rm top}}-h)-(h_{B_{\rm top}}-h_{B}) \right)
\pi^{0}(B,B_{\rm top}) dh \non\\
&\simeq&  \invN M_{0}(h_{B_{\rm top}}-h)  dh 
\simeq \exp \left(-\tau(0)h\right)  dh\non\\
&\simeq& z^{s-1} dz, \label{omega_power}
\een
where $s=\tau(0)$ is the silhouette.
In the last equation we used $N\simeq M_{0}(h_{B_{\rm top}})$ 
where $B_{\rm top}$ is the highest branch point.
In the above equations, we approximated the sums by their mean values 
\eq{Mq} assuming that the contributions of the deviations 
from this mean value vanish 
in the thermodynamics limit $N\rightarrow\infty$, i.e. {\it self-averaging}.
This assumption is valid on trees generated by the RBP because quantities on
sub-trees under different branch points at the same height
are statistically independent from each other.

Then using \eq{omega_power},
we obtain a power law decay in the off-equilibrium limit
\eql{offeq_q}{q(0,t) 
= \int dz \Omega(z) \exp(-zt) 
\sim t^{-\lambda},}
where the exponent $\lambda$ is equal to the fractal dimension or silhouette
of the tree,
\eql{lambda}{\lambda=s=\tau(0).}
The above result is similar to those of the
previously studied hierarchical diffusion models, which also
yield power law decays whose exponents are related with the silhouette
of the trees \cite{HK85}-\cite{BH}.
It is, however,  not surprising because 
the distributions of the equilibrium probabilities in such models
are uniform which is also the case for the present choice of 
the initial condition \eq{initial}.

\subsubsection{Infinite Waiting Time}

The  special case of infinite waiting time $\tw=\infty$ is 
also of interest. In this limit, the system is fully-equilibrated
or aged at $t=0$.
Using $p_{i}(\infty)=\peq_{i}$ and \eq{peq}, we obtain
\ben
\rho_{\infty}(z)
&=&\sum_{B}  \delta(z-\hat{z}(B))
\left \{ \sum_{i} \peq_{i}\pi(L_{i},B)u^{2}(L_{i},B).
\right \} \label{rho_infty}\\
& = & \sum_{B}\delta(z-\hat{z}(B))
\pi(B,B_{\rm top})
\sum_{i} \pi^{2}(L_{i},B)u^{2}(L_{i},B)\non\\
& = &  \sum_{B}\delta(z-\hat{z}(B))
\pi(B,B_{\rm top})\Phi_{B}(z),
\een
where we defined
\ben
\Phi_{B}(z) &\equiv& \delta(z-\hat{z}_{B}) 
\sum_{L_{i} \in {\cal D}(B)} \pi^{2}(L_{i},B)u^{2}(L_{i},B)\\
& = & \frac{\psi_{B}}{1-\psi_{B}}
\delta(z-\hat{z}_{B})\sum_{L_{i} \in {\cal D}_{I}(B)} \pi^{2}(L_{i},B) \non\\
& + & \frac{1-\psi_{B}}{\psi_{B}}
\delta(z-\hat{z}_{B})\sum_{L_{i} \in {\cal D}_{II}(B)} \pi^{2}(L_{i},B).
\een

It is sufficient to consider the scaling property of the
1st term in the last equation. We can rewrite it as,
\eql{tau2}{
\delta(z-\hat{z}_{B})\sum_{L_{i} \in {\cal D}_{I}(B)} \pi^{2}(L_{i},B) 
\sim M_{2}(h) \sim z^{-\tau(2)},}
where we wrote $h\equiv -\log(z)$. In the first equation, we 
evaluated the sum by the mean value \eq{Mq} assuming 
{\it self-averaging} property and used 
\eq{Mq_tauq}. 
Thus $\Phi_{B}(z)$ scales with $z$
as $\Phi_{B}(z) \sim z^{-\tau(2)}$.
In the same way we obtain,
\ben
\sum_{B}\delta(z-\hat{z}_{B})\pi(B,B_{\rm top})dz   &\sim&
M_{1}(h_{B_{\rm top}}-h)  dh\non\\
&\sim&  \mbox{const} \frac{dz}{z}
\een
where we wrote again $h=-\log(z)$ and used  $\tau(1)=0$.
Combining above results we obtain 
\eql{rho_inf}{
\rho_{\infty}(z)dz \sim z^{-\tau(2)-1}dz.}
Using \eq{rho_inf} in \eq{q_rho},
we finally obtain another power law in the fully-equilibrated limit,
\be
q(t+\infty,\infty) \sim t^{-x},
\ee
where $x$ is an exponent defined as,
\eql{x}{x=-\tau(2).}

Let us make some comments on the difference of the two dynamical exponents
$\lambda$ and $x$. Generally, the inequality $\lambda \geq x$ 
and equivalently $\tau(0) \geq -\tau(2)$  hold
as long as $\tau(q)$  decreases monotonically with increasing $q$ and 
concave downward. (Note that $\tau(1)=0$ holds always as we mentioned
before.)
The latter conditions seem to be usually satisfied.
This is certainly the case on the mass exponents 
of trees generated by the RBP whose formula is given in \eq{tauq}.
The equality holds only
when $\pbra=1.0$ (deterministic branching) 
and $F(\psi)=\delta(\psi-0.5)$ (always symmetric partition)
so that the equilibrium probability distribution 
becomes the same 
as the initial non-equilibrium (non-multifractal) distribution.

\subsection{Crossover Behavior : Aging Effect}

Now we consider the case of {\it finite} waiting time, in which
we expect some waiting time effects, i.~e crossover from 
quasi-equilibrium to off-equilibrium behavior.
It is now convenient to introduce another kernel 
$\tilde{\rho}(z,\acute{z})$ such that
\eql{rho_re}{\rho_{\tw}(z) \equiv \int
d\acute{z}\tilde{\rho}(z,\acute{z})\exp(-\acute{z}\tw).}
Then the autocorrelation function \eq{q_rho} 
can be rewritten as
\eql{q_re}{q(t+\tw,\tw)=
\int dz
\int d\acute{z} 
\tilde{\rho}(z,\acute{z})
\exp(-zt)\exp(-\acute{z}\tw).}
From \eq{rho_tw}, \eq{ptw} and \eq{re_G} we obtain the explicit form of the kernel
$\tilde{\rho}(z,\acute{z})$ as
\ben
\tilde{\rho}(z,\acute{z})
&=& \invN \sum_{B}\sum_{\acute{B}}
\delta(z-\hat{z}(B))\delta(\acute{z}-\hat{z}(\acute{B}))\non\\
&\times & \left \{ \sum_{i}\pi(L_{i},B)\pi(L_{i},\acute{B})
u^{2}(L_{i},B)u(L_{i},\acute{B})\sum_{j}u(L_{i},\acute{B}) \right \} .\label{til_rho}
\een
The scaling form of $\tilde{\rho}(z,\acute{z})$
is studied in Appendix C. Here we read the result,
\ben
\tilde{\rho}(z,\acute{z})dz d\acute{z} \sim \left \{
\ba{rc}
z^{-\tau(2)}\disp\frac{dz}{z}\frac{d\acute{z}}{\acute{z}} &
(z > \acute{z})\\\\
\disp \left ( \frac{z}{\acute{z}}\right )^{\tau(0)}
\acute{z}^{-\tau(2)} 
\frac{dz}{z}\frac{d\acute{z}}{\acute{z}} 
& (z < \acute{z}).\label{scale_rho}
\ea \right.
\een
Using \eq{scale_rho} in \eq{q_re}, we finally obtain
\eql{scaleq_udiff}{
q(t+\tw,\tw) \sim
t^{\tau(2)}\tilde{q}_{1}(t/\tw)
+t^{-\tau(0)}\tw^{\tau(0)+\tau(2)}\tilde{q}_{2}(t/\tw),}
where we defined
\ben
\tilde{q}_{1}(s) &\equiv& C_{1} \int dy y^{-\tau(2)-1}\exp(-y)
\int^{y/s}d\acute{y} \acute{y}^{-1}\exp(-\acute{y})\non\\
\tilde{q}_{2}(s) &\equiv&
C_{2} \int dy y^{\tau(0)-1}\exp(-y)
\int_{y/s}d\acute{y}\acute{y}^{-(\tau(0)+\tau(2))-1}\exp(-\acute{y}),
\een
where $C_{1}$ and $C_{2}$ are numerical prefactors. 

From the above results, we find that  
the autocorrelation function obeys the following simple 
$t/\tw$ type scaling,
\be
q(t+\tw,\tw) \sim t^{-x} \tilde{q}(t/\tw),\label{scale1}
\ee
where the scaling function $\tilde{q}(y)$ behaves as
\be
\tilde{q}(y) \sim \left \{ \ba{ll}
\mbox{const} & (y\ll 1)\non\\
y^{x-\lambda} & (y\gg 1), \non\\
\ea \right.\label{scale2}
\ee
with $x=-\tau(2)$ and $\lambda=\tau(0)$.

From the above scaling from, it can be seen that 
the autocorrelation function crossovers from quasi-equilibrium
behavior $t^{-x}$ to off-equilibrium behavior $t^{-\lambda}$
at around $t \sim \tw$.
This crossover behavior appears due to the growth of
the quasi-equilibrium domain in which the probability distribution 
is multifractal while on larger scale, 
it is still non-equilibrium (non-multifractal).
The inequality of the two exponents $\lambda \geq x$
means that
the off-equilibrium decay is {\it faster} than
the quasi-equilibrium decay. which is intuitively satisfactory.

We show in Fig.~4 some  examples of the crossover behavior of 
the autocorrelation function, 
which was obtained using the exact solutions of the
master equation on real samples of random trees.
The random trees are generated
by Monte Carlo method which simulate the
RBP of a) $\pbra=0.10$, $db=1.0$ and $F(\psi)=\delta(\psi-0.2)$ 
and b) $\pbra=0.10$, $db=1.0$ and $F(\psi)=\delta(\psi-0.5)$.
The random average was took over $10^3$ different realizations of
such trees generated by $50$ Monte Carlo steps.
The predicted power law $t^{-\lambda}$
and $t^{-x}$ with a) $\lambda=\tau(0)=0.0953...$ and
$x=-\tau(2)=0.0325...$
(see Fig.~2) 
and b) $\lambda=\tau(0)=0.0953...$ and $x=-\tau(2)=0.0513...$
which are obtained from \eq{tauq}, are also included in the figure.
The curvatures of the curves at lower values of $q$ are due to the 
finite size effects.
In Fig.~5 we show the scaling plot of the data shown in Fig.~4 a).
The curves of different $\tw$  are plotted against $t/\tw$ and 
shifted vertically so as to converge to a master curve.
One can well see that the data are consistent with the predicted scaling
laws \eq{scale1} and \eq{scale2}. 

\section{Discussions}

We have studied aging effects in 
a simple exactly solvable model of hierarchical diffusion. 
We considered the case that equilibrium probability distribution 
has multifractality. A specific way to generate such trees  by 
randomly branching processes (RBP) are introduced for demonstrations.
Aging processes after temperature quenches appear 
as the growth of aged sub-trees
in which the probability distribution is in quasi-equilibrium and
multifractal. In the thermodynamics limit,
the height of the tree becomes infinite and the 
true equilibrium cannot be attained in any large but finite 
waiting time $\tw$. 
Consequently, the waiting time dependence persists for the whole 
range of $\tw$ except for $\tw=\infty$, 
i.~e the ergodicity is weakly broken \cite{B92}.
We found that these properties clearly reflect in 
the survival-return probability and brings about 
the characteristic crossover from quasi-equilibrium
behavior to off-equilibrium behavior.

Let us make some comments on the robustness of the scaling properties
of the autocorrelation function we obtained in our exactly solvable model.
Note that there are other possible choices of the transition matrix 
other than  our present choice,
which describe hierarchical diffusions and 
endures the detailed balance condition \eq{d_balance}. 
For instance, one may define another transition matrix by replacing
$\pi(L_{i},A_{n})$ in \eq{Wij} by $\pi(L_{j},A_{n})^{-1}$.
One can also construct transition matrices considering that
transitions between a pair of leaves occur only over their lowest common
ancestor. We investigated the solutions of the master equations with these
alternative transition matrices
on random trees generated by the RBP by numerical diagonalizing the
transition matricies. 
Interestingly enough, we found that the scaling behaviors of the 
autocorrelation function appears essentially the same as that 
of the exactly solvable one presented in this paper and
one only needs some renormalization of the global unit of time.
This fact implies that the scaling properties are robust to some extent
against minor changes of the model.

It is interesting to note that the crossover behavior from 
{\it quasi-equilibrium} behavior ($t\ll\tw$) 
to {\it off-equilibrium} behavior ($t\gg\tw$) 
obtained in the preset toy model, is very similar to that 
observed in the relaxational dynamics of some
micro-scopic models of random systems such as 
3D spin-glass model \cite{R93} and 
1+1 dimensional directed polymer in random media \cite{Y}.
In the latter models, the autocorrelation functions obey
$t/\tw$ scaling law with two power law decays $t^{-x}$ at
the quasi-equilibrium regime and $t^{-\lambda}$ at the 
off-equilibrium regime, which is what we found 
in the present phenomenological toy model.
Thus it is tempting to speculate that our present toy model 
will provide a  clue to understand the link 
between phenomenological pictures based on hierarchical diffusion 
and the glassy dynamics of realistic systems \cite{remark}.

Acknowledgment: The author would like to thank sincerely 
Prof.~Takayama for valuable discussions and critical reading
of the manuscript. 
The communications with Prof. ~Bouchaud and Prof.~Sibani 
are gratefully acknowledged.
He would also like to thank Prof.~Nemoto,  Prof.~Arimitsu
and  Dr.~Hukusima for stimulating discussions and valuable comments.
This work was supported by Grand-in-Aid for Scientific Research 
from the Ministry of Education, Science and Culture, Japan.
The author was supported by Fellowships of the Japan Society
for the Promotion of Science for Japanese Junior Scientists.

\newpage
\appendix
\section{The Umbrella Set}

In this appendix we show that the umbrella set 
defined in \eq{umb}, \eq{u_d} and \eq{z_umb} are the true eigen states
of the matrix $\tilde{\bf \Gamma}$ defined in \eq{tildeG}.
We prove this by checking if the umbrella set correctly
reproduce the original transition matrix defined in \eq{Wij}.

Using \eq{umb} and \eq{z_umb} in \eq{Gamma_re} we obtain 
\ben
 -{\bf \Gamma}_{i\neq j} & = & 
-\sum_{B}\pi(L_{i},B)u(L_{i},B)\hat{z}(B)u(L_{j},B)\non\\
&=& \pi(L_{i},A(L_{i},L_{j}))\exp(-h_{A(L_{i},L_{j})})\non\\
&& - \sum_{B \in {\cal A}(A(L_{i},L_{j}))}
\pi(L_{i},B)\exp(-h_{B})u^{2}(L_{j},B)\non\\
&=& \pi(L_{i},A_{0})\exp(-h_{A_{0}})
- \sum_{n=1}^{M}\pi(L_{i},A_{n})\exp(-h_{A_{n}})u^{2}(L_{j},A_{n})\non
\een
where we defined $A_{0}=A(L_{i},L_{j})$ and $A_{1},A_{2},\ldots,A_{M-2}$,
$A_{M-1}=B_{\rm top}$ and $A_{M}=B_{\rm ceiling}$.

We further rewrite the right hand side of the last equation
as the following.
The factor $u(L_{j},A_{n})$ in the last equation can be replaced 
by $u(A_{n-1},A_{n})$ due to the definition \eq{u_d}.
And the factor $\pi(L_{i},A_{n})$ can be decomposed as 
$\pi(L_{i},A_{n-1})\pi(A_{n-1},A_{n})$.
Then we can use the identity
$\pi(A_{n-1},A_{n})u^{2}(A_{n-1},A_{n})=1-\pi(A_{n-1},A_{n})$, 
which follows from \eq{u_d} and \eq{rw}. 
Then we obtain
\ben
\mbox{r. h. s.} &=& \pi(L_{i},A_{0})\exp(-h_{A_{0}})\non\\
&\ & -\sum_{n=1}^{M}\pi(L_{i},A_{n-1})\left(1-\pi(A_{n-1},A_{n})\right)
\exp(-h_{A_{n}})\non\\
&=&  \disp\sum_{n=0}^{M-1}
\left[ \exp(-h_{A_{n}})-\exp(-h_{A_{n+1}}) \right]\pi(L_{i},A_{n})
\een
where we used $h_{A_{M}}=h_{B_{\rm ceiling}}=\infty$ in the last equation.
Then using the relation \eq{Gamma}, we see that the off-diagonal elements
of the transition probability ${\bf W}_{i\neq j}$ defined in
equation \eq{Wij} is correctly reproduced by the umbrella set.

Next we check 
if the matrix ${\bf \Gamma}$ written in terms of the umbrella set,
properly conserve the total probability. 
Taking the sum over $i$ of both sides of \eq{Gamma_re} we obtain
\ben
\sum_{i}{\bf \Gamma}_{ij}&=&\sum_{B}
\left(\sum_{i}\pi(L_{i},B)u(L_{i},B)\right)\hat{z}(B)u(L_{j},B)\non\\
&=& 0 \label{check_Ga}
\een
The last equation follows from the following.
For $B=B_{\rm ceiling}$, the contribution is $0$ since 
$\hat{z}(B_{\rm ceiling})=0$. And for $B\neq B_{\rm ceiling}$, 
one finds again zero contributions using \eq{umb},
\ben
\sum_{i}\pi(L_{i},B)u(L_{i},B) &=& 
\disp\sqrt\frac{\psi_{B}}{1-\psi_{B}}\sum_{L_{i}\in {\cal D}_{I}(B)}
\pi(L_{i},B)
-\disp\sqrt\frac{1-\psi_{B}}{\psi_{B}}\sum_{L_{i}\in {\cal D}_{II}(B)}
\pi(L_{i},B)\non\\
&=&\disp\sqrt\frac{\psi_{B}}{1-\psi_{B}}(1-\psi_{B})
-\disp\sqrt\frac{1-\psi_{B}}{\psi_{B}}\psi_{B}=0\non.
\een
Thus we obtain the last equation of \eq{check_Ga}.

\section{Mass Exponents on Random Trees}

In this appendix we study 
the scaling property of the $q$-th moment of the probability distribution 
on the random trees generated by the randomly branching process (RBP)
and derive the formula \eq{tauq}. 

Let us denote the probability 
that the q-th moment $M_{q}(h)$ of a random tree of size $h=m db$ 
takes value $x$ as $\omega_{q}(m,x)$. 
Considering that larger
trees can be constructed by smaller sub-trees,
we obtain the following recursion relation for
$\omega_{q}(m,x)$,
\ben
\omega_{q}(m+1,x) &=& \pbra\int dy_{1}dy_{2}
\int_{0}^{1} d\psi F(\psi) \omega_{q}(m,y_{1})\omega_{q}(m,y_{2})\non\\
&\times &\delta(\psi^{q}y_{1}+(1-\psi)^{q}y_{2}-x) 
+ (1-\pbra)\omega_{q}(m,x)\label{omega_rec}.
\een

In order to solve this integral equation, it is convenient 
to introduce a generating function defined as
\eql{zq}{Z_{q}(u,m)\equiv\int dx \exp(ux)\omega_{q}(m,x).}
The expectation value of the q-the moment can be obtained as
\eql{expect_Mq}{M_{q}(h) 
\simeq <x>_{q,m}\equiv\int dx x \omega_{q}(m,x)
=\left. \frac{\partial}{\partial u}Z_{q}(u,m)\right|_{u=0}.}
Multiplying $\exp(ux)$ on both sides of \eq{omega_rec} and
integrating over $x$, we obtain the recursion relation for $Z_{q}(u,M)$
\ben
Z_{q}(u,m+1)&=&\pbra \int_{0}^{1} d\psi F(\psi)
Z_{q}(u\psi^{q},m)Z_{q}(u(1-\psi)^{q},m) \non\\
&+& (1-\pbra)Z_{q}(u,m).
\een
Then we obtain the recursion relation for $<x>_{q,m}$,
\eql{xq_re}{<x>_{q,M+1}=
\left[ \pbra \int_{0}^{1} d\psi F(\psi)
\{\psi^{q}+(1-\psi)^{q}\} +(1-\pbra) \right ] <x>_{q,M}.}
Solving the last equation with  $<x>_{q,1}=1$, we get
\eql{xq_result}{M_{q}(h) \sim <x>_{q,M}=\exp(\tau(q)h),}
where the mass exponent $\tau(q)$ is obtained  as
\eql{tauq_result}{\tau(q)=
db^{-1}\log \left [ \pbra \disp\int_{0}^{1}d\psi F(\psi)
\left\{\psi^{q}+(1-\psi)^{q} \right\}+(1-\pbra) \right ].}

In order to further investigate the multifractal properties,
it is convenient to introduce 
the {\it exponent of singularity} $\alpha$ defined as
\be
\pi (B,\acute{B})\equiv \exp \left[-\alpha(B,\acute{B})
(h_{B}-h_{\acute{B}})\right]. \label{mf_hdiff}
\ee
Then one can obtain the distribution of $\alpha$ or $f(\alpha)$ spectrum 
using the well known procedure \cite{multifractal} and discuss {\it curdling}.
However we don't discuss about it here \cite{thesis}.

\section{Scaling form of Kernel}

In this appendix we study the scaling property of the kernel
$\tilde{\rho}(z,\acute{z})$ with respect to $z$ and $\acute{z}$.
Its explicit form \eq{til_rho} is
\ben
\tilde{\rho}(z,\acute{z})
&=& \invN \sum_{B}\sum_{\acute{B}}
\delta(z-\hat{z}(B))\delta(\acute{z}-\hat{z}(\acute{B}))\non\\
&\times & \left \{ \sum_{i}\pi(L_{i},B)\pi(L_{i},\acute{B})
u^{2}(L_{i},B)u(L_{i},\acute{B})\sum_{j}u(L_{j},\acute{B}) 
\right \}. \label{scale_re}
\een
Note that, from the definitions \eq{u_d}, 
the factor $u^{2}(L_{i},B)u(L_{i},\acute{B})$ is non-zero only
when the leaf $L_{i}$ is under both $B$ and $\acute{B}$.

At first we consider the case $z > \acute{z}$.
In this case, the terms which survive in the sum \eq{scale_re}
are those in which $\acute{B}$ is an
ancestor of $B$ and $B$ is an ancestor of $L_{i}$. 
Let us introduce 
$h\equiv -\log(z)$ and
$\acute{h} \equiv -\log(\acute{z})$.
Then we obtain 
\ben
\tilde{\rho}(z,\acute{z})dz d\acute{z} 
\ \ \ \ \ ( z > \acute{z})
&=& \invN\sum_{\acute{B} \in {\cal D}(B_{\rm top})}
\delta({\acute{z}-\hat{z}(\acute{B})})\pi^{0}(\acute{B},B_{\rm top})
\non\\
&\times& \sum_{L{j} \in {\cal D}(\acute{B})} 
\delta(\acute{z}-\hat{z}(B))
\pi^{0}(L_{j},\acute{B})u(L_{j},\acute{B}) \non\\
&\times& \sum_{B\in {\cal D}(\acute{B})}
\delta({z-\hat{z}(B)})\delta({\acute{z}-\hat{z}(\acute{B})})
\pi^{1}(B,\acute{B})u(B,\acute{B})\non\\
&\times& \sum_{L_{i}\in {\cal D}(B)}
\delta(z-\hat{z}(B))\pi^{2}(L_{i},B)
u^{2}(L_{i},B)  dz d\acute{z}\non\\
& \sim& \invN M_{0}(h_{B_{\rm top}}-\acute{h})M_{0}(\acute{h})
M_{1}(\acute{h}-h)M_{2}(h) dhd\acute{h} \non\\
&\sim& z^{-\tau(2)} 
\frac{dz}{z} \frac{d\acute{z}}{\acute{z}}
\een
where we evaluated the sums by their mean values \eq{Mq},
assuming {\it self-averaging} property,
and used \eq{tauq} and $N\simeq M_{0}(h_{B_{\rm top}})$.

The other case $z < \acute{z}$ can be analyzed in the same way.
Considering that the terms which survive in the sum \eq{scale_re}
are those in which $B$ is an
ancestor of $\acute{B}$ and $\acute{B}$ is an ancestor of $L_{i}$, 
we obtain
\ben
\tilde{\rho}(z,\acute{z})dz d\acute{z} 
\ \ \ \ \ ( z < \acute{z})
&=& \invN\sum_{B \in {\cal D}(B_{\rm top})}
\delta({z-\hat{z}(B)})\pi^{0}(B,B_{\rm top})
\non\\
&\times& \sum_{\acute{B}\in {\cal D}(B)}
\delta({z-\hat{z}(B)})\delta({\acute{z}-\hat{z}(\acute{B})})
\pi^{1}(\acute{B},B)u^{2}(\acute{B},B)\non\\
&\times& \sum_{L_{i} \in {\cal D}(\acute{B})} 
\delta(\acute{z}-\hat{z}(B))
\pi^{2}(L_{i},\acute{B})u(L_{i},\acute{B}) \non\\
&\times& \sum_{L_{j}\in {\cal D}(\acute{B})}
\delta(z-\hat{z}(B))\pi^{0}(L_{i},\acute{B})
u(L_{i},\acute{B})  dz d\acute{z}\non\\
&\sim& \invN M_{0}(h_{B_{\rm top}}-h)M_{1}(h-\acute{h})
M_{2}(\acute{h})M_{0}(\acute{h}) dhd\acute{h} \non\\
&\sim& \disp\left ( \frac{z}{\acute{z}} \right )^{\tau(0)}
\acute{z}^{-\tau(2)} 
\frac{dz}{z} \frac{d\acute{z}}{\acute{z}}
\een
where we used again $N\simeq M_{0}(h_{B_{\rm top}})$.
Combining above results, we obtain 
\eq{scale_rho}.

\setlength\baselineskip{7mm}

\newpage
\noindent
{\bf \large FIGURE CAPTIONS}

\vspace*{3mm}\noindent
Fig.~1 Hierarchical organization of {\it states}.
       This example is 
       generated by a RBP explained in section \ref{RBP}
       ($\pbra=0.10$, $db=1.0$, $F(\psi)=\delta(\psi-0.2)$, $20$ MCS.)
        All the branches on the left side
        have weight $1-\psi=0.8$ while those on the right side
        have weight $\psi=0.2$ as indicated in the figure.

\vspace*{3mm}\noindent
Fig.~2  The non-linear behavior of the mass exponent $\tau(q)$ v.s. $q$:
        the curve is obtained using \eq{tauq} for
        RBP of $\pbra=0.10$, $db=1.0$ and $F(\psi)=\delta(\psi-0.2)$.
        The two important values of $\tau(q)$  at 
        $q=0$ and $q=2$ are indicated by the arrows for later
        reference.

\vspace*{3mm}\noindent
Fig.~3   Growth of local equilibrium domain  with increasing 
         waiting time $\tw$ : 
         the plot of $r_{i}(\tw)\equiv p_{i}(\tw)/\peq_{i}$ 
         vs. $\tw$ on the random tree shown in Fig.~2.

\vspace*{3mm}\noindent
Fig.~4   The crossover behavior of $q(\tw+t,\tw)$
at different waiting times $\tw$ on two specific kinds of 
random trees a) and b) (see text) .
The waiting time $\tw$ varies as 
$\tw=0, 10, 10^{2}, \ldots, 10^{14},\infty$
from the lowest  to the top curve. 

\vspace*{3mm}\noindent
Fig.~5   The $t/\tw$ scaling plot of $q(\tw+t,\tw)$: the data of
         of different $\tw$ presented in Fig.~5 a)
         except $\tw=0$ and $\infty$, are used in this plot.
         The vertical scale is arbitrary.

\pagestyle{empty}
\noindent{\large FIGURES}

\newpage
\hbox to \textwidth{
\vtop{
\hsize=15cm
\centerline{
        \epsfxsize=10.0cm
        \epsfysize=10.0cm
        \epsfbox{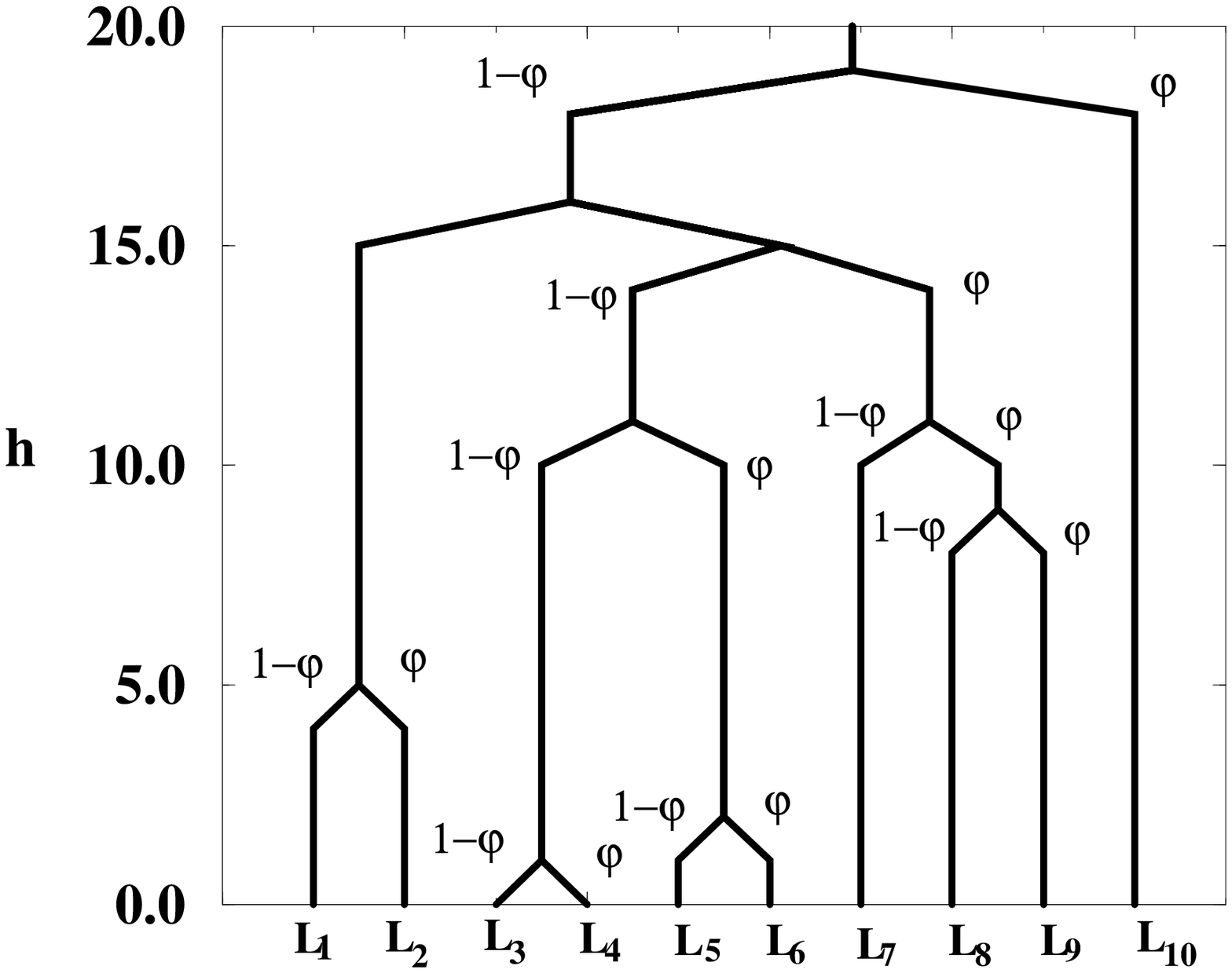}
}
\noindent
Fig.~1 
}}

\newpage
\vspace*{-2cm}
\hbox to \textwidth{
\vtop{
\hsize=15cm
\centerline{
        \epsfxsize=10cm
        \epsfysize=8cm
        \epsfbox{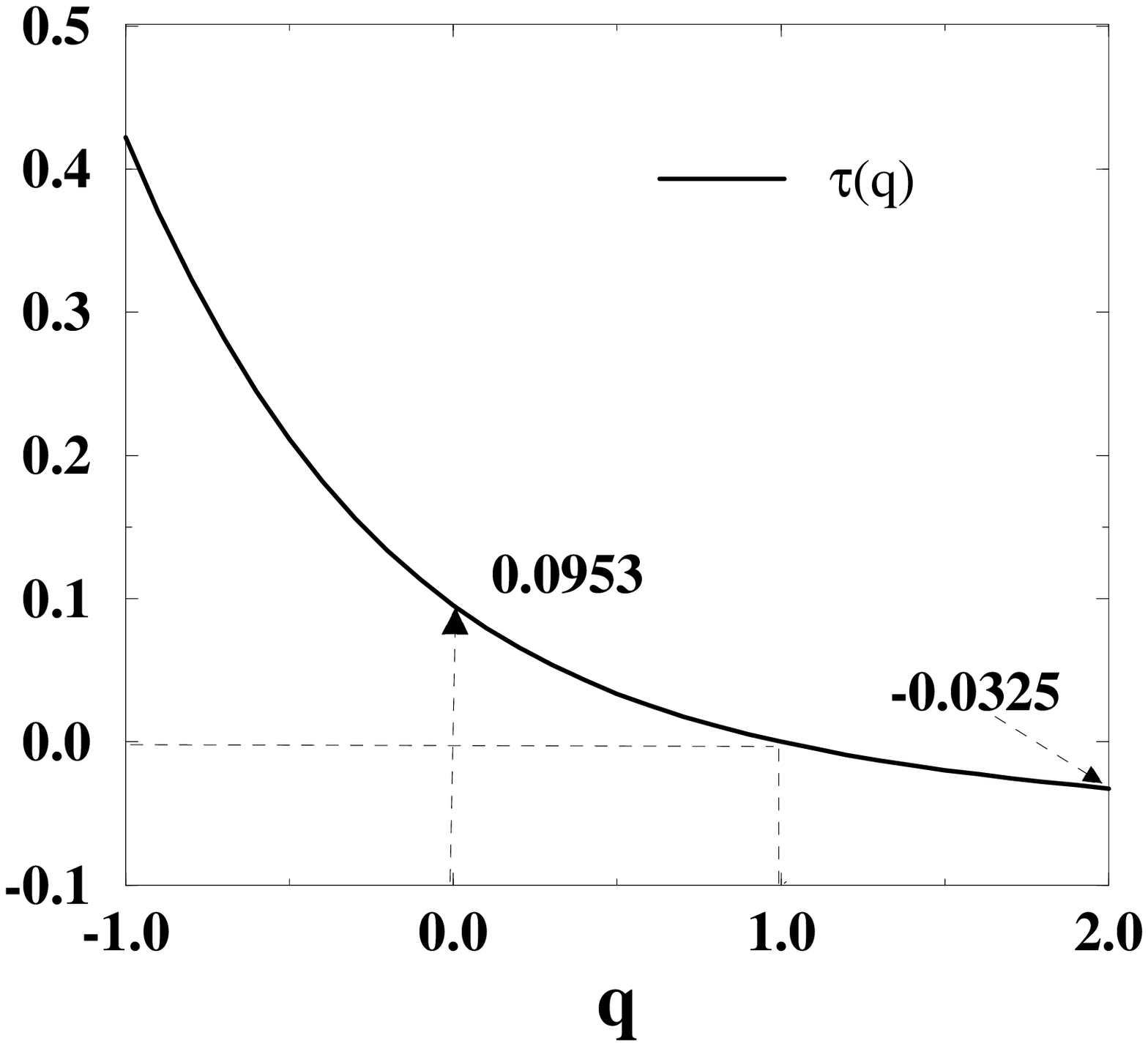}
}
\vspace*{5mm}
\noindent
Fig.~2 
}}

\newpage
\vspace*{1cm}
\hbox to \textwidth{
\vtop{
\hsize=15cm
\centerline{
        \epsfxsize=10cm
        \epsfysize=10cm
        \epsfbox{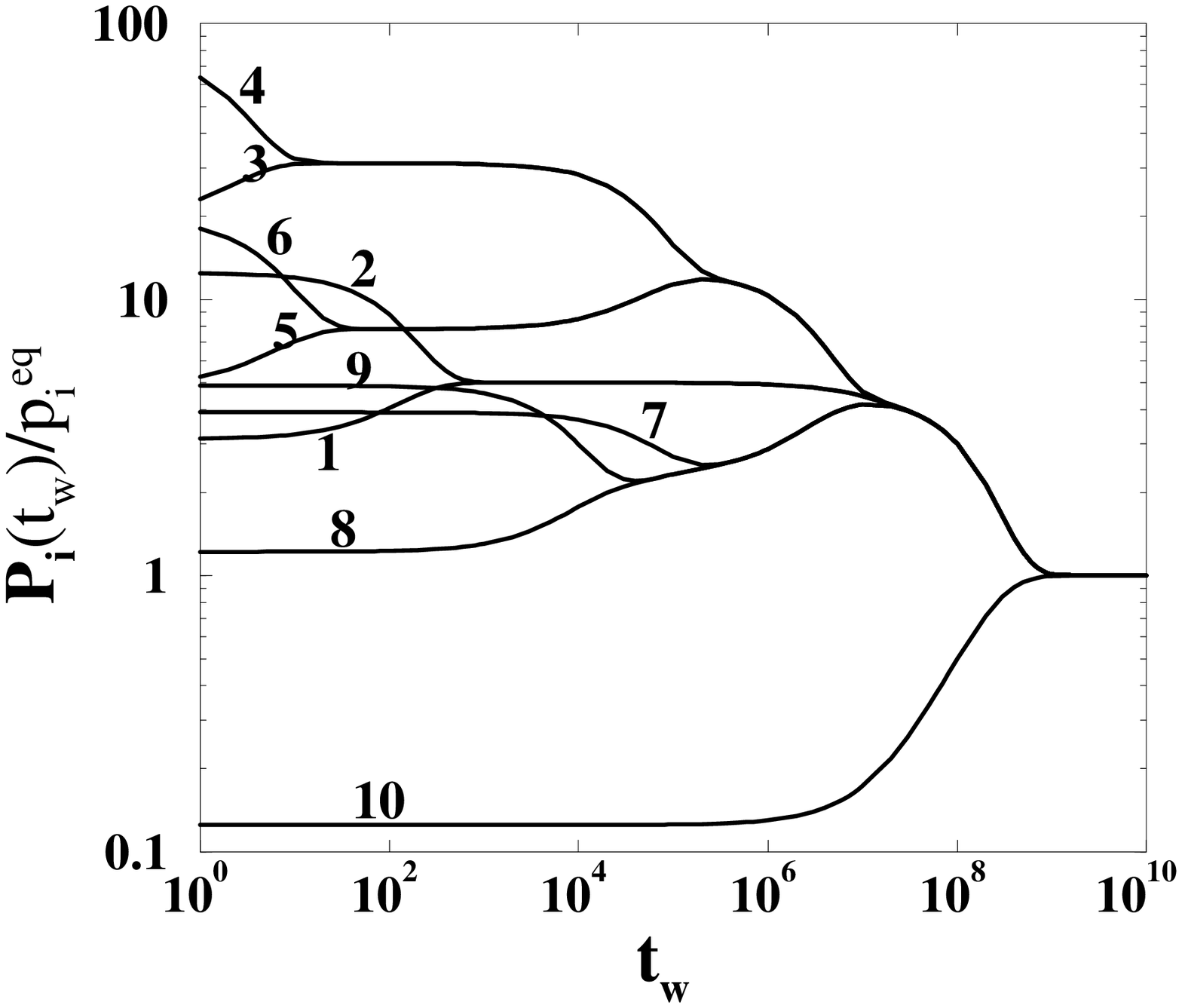}
}
\vspace*{5mm}
\noindent
Fig.~3  
}}

\newpage
\hbox to \textwidth{
\vtop{
\hsize=15cm
\centerline{
        \epsfxsize=9cm
        \epsfysize=7cm
        \epsfbox{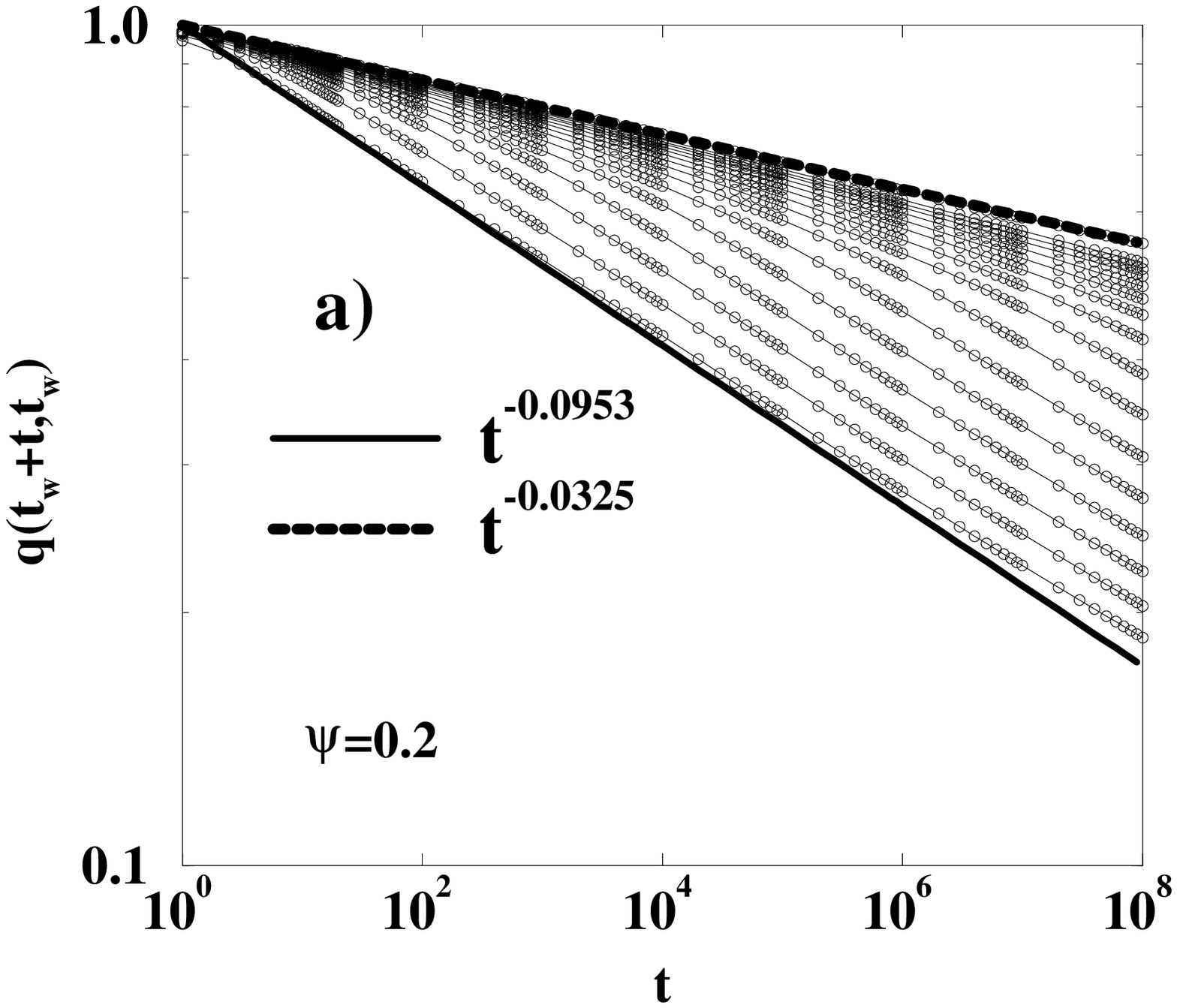}
}
\vspace*{5mm}
\centerline{
        \epsfxsize=9cm
        \epsfysize=7cm
        \epsfbox{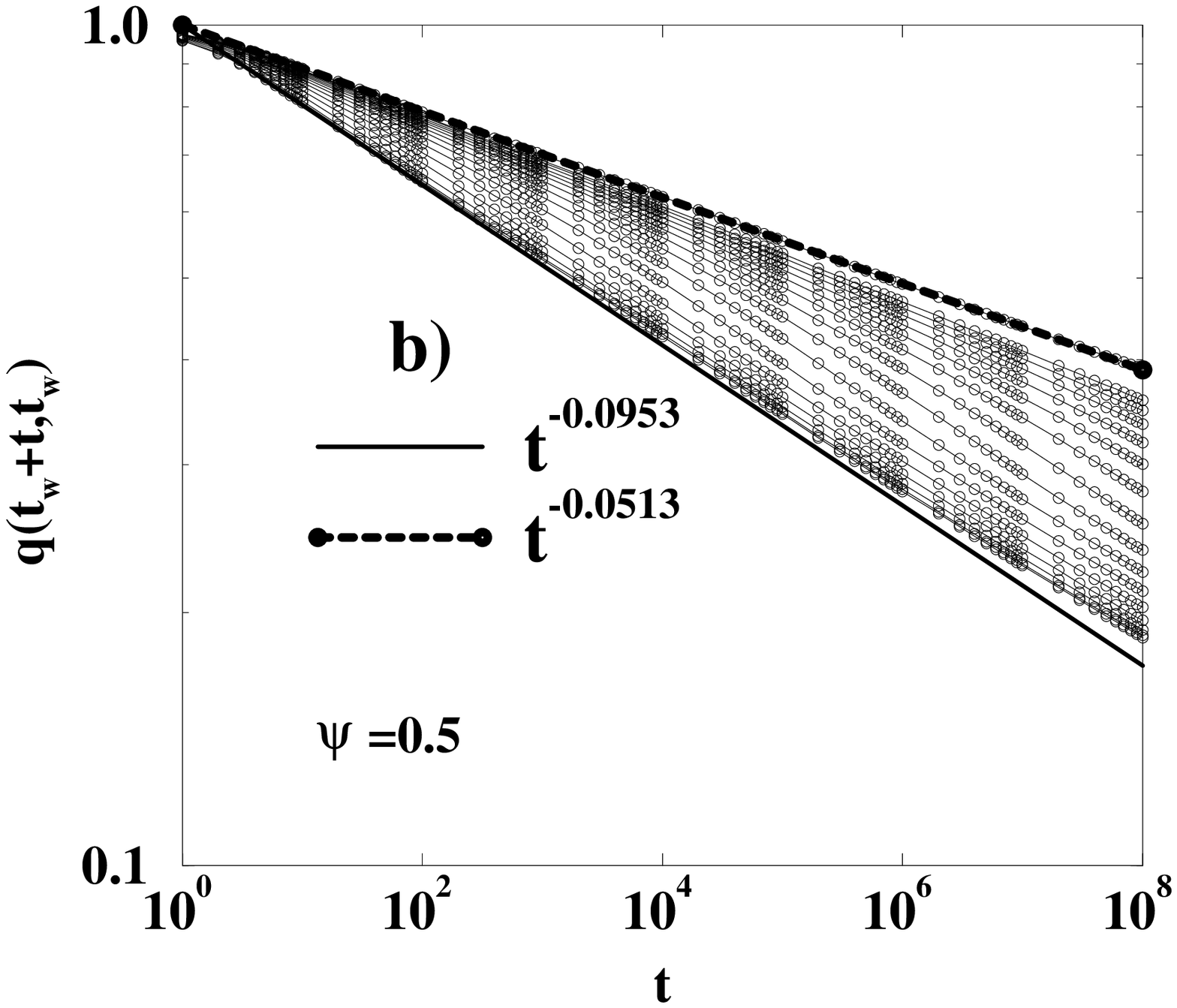}
}
\vspace*{5mm}
\noindent
Fig.~4 
}}

\newpage
\hbox to \textwidth{
\vtop{
\hsize=15cm
\centerline{
        \epsfxsize=10cm
        \epsfysize=10cm
        \epsfbox{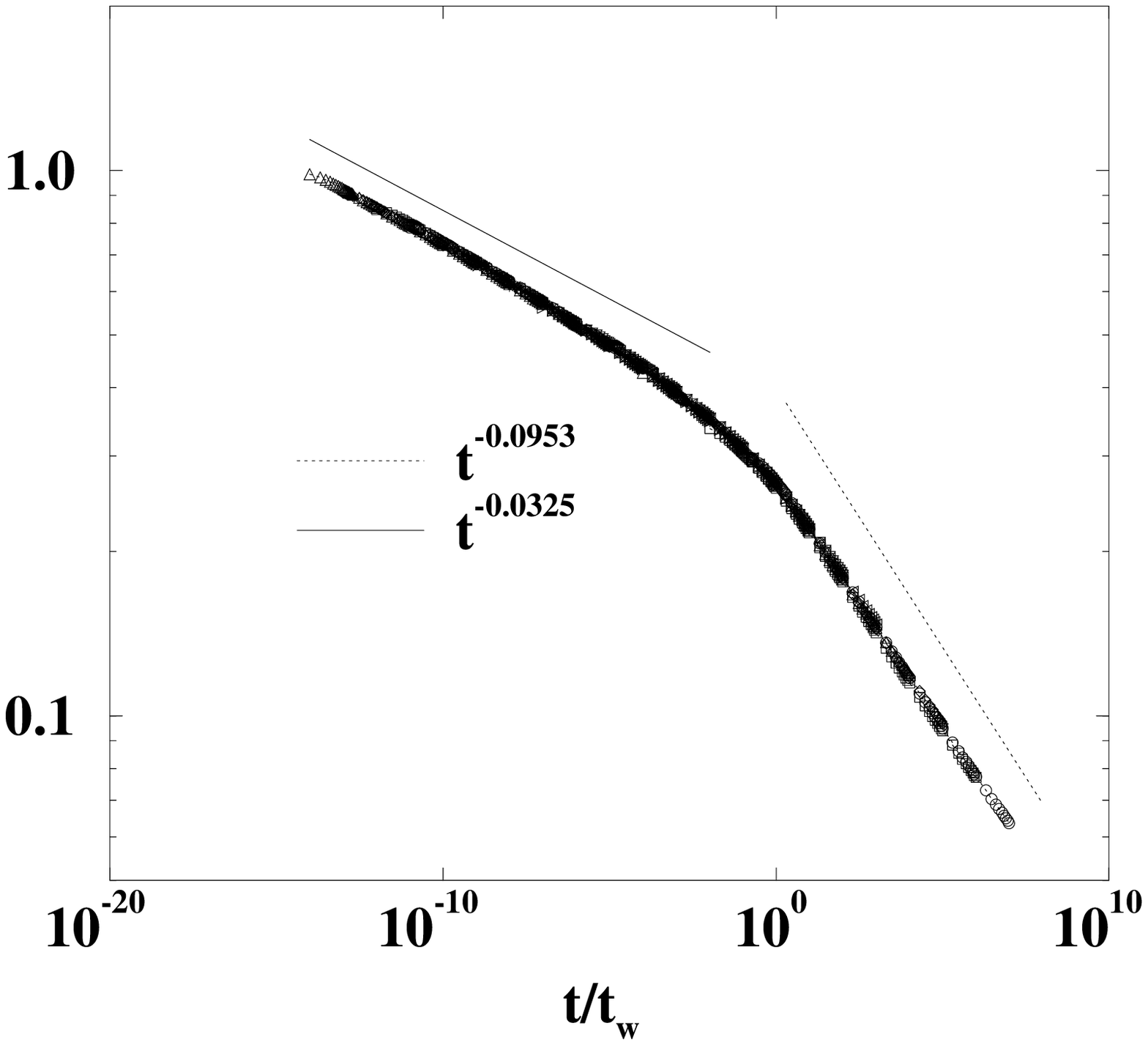}
}
\vspace*{5mm}
\noindent
Fig.~5 
}}

\newpage
\noindent
{\bf \large LIST OF CHANGES MADE IN THIS REVISED VERSION}
\begin{itemize}

\item The subsection which was titled 'Distribution of Relaxation
        Times' in the original version, 
        which was placed just after subsection \ref{srp},
        was removed and the necessary part of it is now
        included in subsection \ref{zero}.

\item The figure 1 of the original version, which was intended
to explain the idea of hierarchical decomposition of the phase space,
was removed. This is because figure 2 in the original version (figure
1 of the revised version), which shows an example of
random trees generated by RBP, can be used to explain the idea.

\item The list of references were updated and some additional
        references ( the last three of it) are added.

\item Minor changes such as corrections of miss typing 
        and refinements of some sentences are made.

\end{itemize}


\newpage
\begin{thebibliography}{99}


\bibitem{RPRM} 
E. Vincent, J. Hamman, and M. Ocio, in 
{\it Recent Progress in Random Magnets} (World Scientific,
Singapore,1992) and references there in.

\bibitem{STRUIK} L.~C.~E Struik, 1978 {\it Physical Aging in Amorphous
Polymers and Other Materials} Elsevier.


\bibitem{SH} P. Sibani and K. H. Hoffmann 1989, 
Phys.\ Rev.\ Lett.\ {\bf 63} 2853,
1990  Z. Phys. {\bf B 80}, 429.

\bibitem{HS88} K. Hoffmann and P. Sibani 1988, Phy. Rev. {\bf A 38} 4261.
\bibitem{HK85} B. A. Huberman and M. Kerszberg 1985, J. Phys. {\bf A 18} L331.
\bibitem{S85} M. Shreckenberg 1985, Z. Phys. {\bf B 60} 483.
\bibitem{OS} A. T. Ogielski and D. L. Stein 1985, 
Phys. Rev. Lett. {\bf 55} 1634.
\bibitem{BH} C. P. Bachas and B. A. Huberman 1987, J. Phys. {\bf A 20} 4995. 

\bibitem{Nemo88} K. Nemoto, 
in {\it Cooperative Dynamics
in Complex Physical Systems}, Hajime Takayama eds. (Springer-Verlag, 1988),
has considered to mimic the dynamics of the SK
model at low temperatures by a hierarchical diffusion model. 
He constructed random trees using information
of the metastable states of the model, in which 
both the heights of the branch points and the statistical weights
of the leaves have randomness. The present work is 
in part inspired by his work.
\bibitem{Pal} R. G. Palmer 1982, Adv. In. Phys. {\bf vol. 31} 669
and 1987, in {\it Heidelberg Colloquium on Glassy Dynamics}
J. L. van Hemmen and I. Morgenstern eds. Lecture Notes in Physics
275 (Springer-Verlag ).


\bibitem{SS94} P.~Sibani and P.~Schriver 1994, Phys. Rev. {\bf B 49} 6667.
see also P.~Sibani, J.~C.~Sch\"{o}n, P.~Salamon, and J.~O.~Anderson
1993, Europhys. Lett. {\bf 22}, 479 .


\bibitem{B92} J.~P.~Bouchaud 1992, J. Phys. (France) {\bf 2} 1705.


\bibitem{R93} Heiko Rieger 1993, J. Phys. {\bf A 26}, L615
and 1995 in {\it Annual Reviews of Computational Physics $II$} ed. D. Stauffer
(World Scientific, Singapore).

\bibitem{Y} Hajime Yoshino 1996, J. Phys. {\bf A 29}, 1421.


\bibitem{Mandel} B.B. Mandelbrot , {\it The Fractal Geometry of Nature}
(Freeman, San Francisco 1982). 
\bibitem{multifractal} J. Feder, {\it Fractals} (Plenum, New York, 1988)

\bibitem{thesis} Hajime Yoshino 1996, Ph. D. thesis.


\bibitem{remark} 
In the case of SK model, which is a mean-field spin-glass model,
it is knowm \cite{MPSTV} 
that the participation ratio $\sum_{\alpha} W_{\alpha}^2$ where
$W_{\alpha}$ is the equilibrium statistical weight of a pure state $\alpha$,
is non-zero. The latter means that the total statistical 
weight is dominated by only a few pure states. 
Contrary, in the case of random trees generated by RBP,
the participation ratio goes to zero as the height of the tree $h$
becomes infinitely high because $\tau(2)$ is negative. 
Thus in the case of RBP trees, the survival-return 
probability goes to zero in the limit $t \rightarrow \infty$ 
even after the limit $\tw \rightarrow \infty$ is took,
while it migth be non-zero in the case of SK model.
I thank J.~P.~Bouchaud for pointing this out \cite{JP}.

\bibitem{JP} J.~P.~Bouchaud, private communication.
\bibitem{MPSTV} M. M\'{e}zart, G. Parisi, N. Sourlas, G. Toulouse and
M. Virasoro 1984, Phys. Rev. Lett. {\bf 52} 1156 and 
J. Physique {\bf 45}  843.

\end{thebibliography}
\end{document}